\documentclass[review]{elsarticle}
\usepackage{amsmath}
\usepackage{amssymb}

\newcommand{\beq}{\begin{center}\begin{equation}}
\newcommand{\eeq}{\end{equation}\end{center}}

\begin{document}
\begin{frontmatter}

\title{An Investigation of Machine Learning Methods Applied to Structure Prediction in Condensed Matter}

\author{William J. Brouwer\footnote{Corresponding author, email address : wjb19@psu.edu},$^a$ James D. Kubicki,$^b$ Jorge O. Sofo,$^c$ C. Lee Giles$^d$}
\address{$^{a}$Research Computing and Cyberinfrastructure, $^{b}$Department of Geosciences, $^{c}$Department of Physics, $^{d}$Information Science and Technology, The Pennsylvania State University}

\begin{abstract}
Materials characterization remains a significant, time-consuming undertaking. Generally speaking, spectroscopic techniques are used in conjunction with empirical and 
ab-initio calculations in order to elucidate structure. These experimental and computational methods typically require significant human input and interpretation, particularly with regards to novel materials. Recently, the application of data mining and machine learning to problems in material science have shown great promise in reducing this overhead~\cite{saad}. In the work presented 
here, several aspects of machine learning are explored with regards to characterizing a model material, titania, using solid state Nuclear Magnetic Resonance (NMR). Specifically, a large dataset is 
generated, corresponding to NMR $^{47}$Ti spectra, using ab-initio calculations for generated TiO$_2$ structures. Principal Components Analysis (PCA) reveals that input spectra may be compressed by more
than 90\%, before being used for subsequent machine learning. Two key methods are used to learn the complex mapping between structural details and input NMR spectra, demonstrating excellent accuracy 
when presented with test sample spectra.
This work compares Support Vector Regression (SVR) and Artificial Neural Networks (ANNs), as one step towards the construction of an expert system for solid state materials characterization.
\end{abstract}

\end{frontmatter}

\section{Introduction}
Structure characterization is one of many interesting problems in material science, the latter recently garnering significant attention in the form of the Materials Genome Initiative~\cite{gen,sci_gen}, which ultimately seeks to understand the atomistic blueprint for key materials. Researchers across various scientific disciplines seek to develop structural models for condensed and molecular systems. The modeling process revolves around the gradual refinement 
of assumptions, through comparison of experimental and computational results. 
A critical experimental technique used in material science is solid-state NMR, a method that provides great insight into 
chemical order over Angstrom length scales, an important spectroscopic tool used in key discoveries~\cite{pet}. However, interpretation of spectra for new and complex solid-state materials is difficult, often requiring experiments on model compounds in order to derive empirical relationships, for interpretation of the system under study~\cite{emp}, overall 
a time-consuming process. Spectra must also generally be simulated and fit in order to extract parameters that correspond to structural features. 

This process of simulation and fitting 
is common to many experimental techniques, including X-ray spectroscopy. Similarly, working forward from structural models in order to produce measurable experimental quantities 
calculated from first principles is computationally demanding. Thus, the process of structure determination has significant impediments, slowing the time to discovery significantly. The present work explores the introduction of machine learning 
to materials characterization, specifically to quantify structural distortions in simple oxides, which has relevance to characterizing more complex oxides including Relaxor Ferroelectrics~\cite{hoat}.

The application of machine learning is common to several disciplines, for instance, recent work has been devoted to creating a method to predict the outcome of chemical reactions 
in organic chemistry, provided with input reactants and conditions\cite{exp1}. Expert systems have been developed for structure determination of organic molecules, from acquired 
crystallographic and NMR data \cite{exp2,exp3}. Similarly, a variety of approaches have been devised for determining protein structure from 2D NMR experiments\cite{exp4}. Also 
within structural biology, machine learning has been applied to predict protein-ligand binding affinity\cite{exp5}. 

These chemical or molecular examples stand in distinction to the
 characterization of condensed or solid-state materials, where methods of calculating fundamental quantities and spectroscopic observations present different challenges. For example, although
protein structure is complicated and nuanced, 2D and 3D NMR spectra of proteins are generally composed of well-resolved lines that have direct correlations with structural details such as 
bond lengths and chemical species. On the other hand, solid-state spectra for condensed materials generally suffer from line broadening mechanisms to be discussed shortly that degrade spectral 
resolution and make interpretation difficult. This degradation in resolution is made worse by local disorder in bonding environments, for example in glasses and solid solutions. Similar 
difficulties pervade other spectroscopic techniques including X-Ray diffraction. 
With regards to first-principle calculations of spectroscopic quantities, molecular systems can prove formidable 
but manageable computationally. Gaussian-based orbitals have been used for many decades in the solution of the Schrodinger equation for molecular systems~\cite{fri}, and in conjunction with 
approximations for electronic exchange and correlation effects, provide accurate values for a wide variety of measurable quantities. On the other hand, calculations of electronic structure in 
extended systems require the use of periodic boundary conditions and plane wave orbitals for the electronic states, which is much less computationally tractable, for reasons to be discussed 
shortly. Nonetheless, great strides have been made in the development and use of ab-initio calculations in solid-state material science. The use of Density Functional Theory (DFT)~\cite{dft} in 
particular has increased dramatically over the last decade, permitting scientists to evaluate potentially useful materials computationally~\cite{cur1,cur2}, without the need for costly 
synthesis and spectroscopy. As a computational tool, DFT also allows for the study and structure determination of inaccessible materials eg., the inner core of terrestrial planets~\cite{fe}.

\subsection{Contribution}

Machine learning is generally used in order to derive useful information and relationships, particularly for large datasets. 
Machine learning exists in many forms, although all methods may generally be regarded as being either unsupervised or supervised in nature. Supervised methods are those that take a number 
of data examples during a training phase, where the input and output spaces for the data can be widely varying in size and nature. During the training phase, a mapping between the 
two spaces is determined and represented in a model germane to the method used, such that when presented with new data, predicted output is returned for a 
given input. Predictions might be binary by nature i.e., classification or numerical i.e., regression, or a combination. Attention in this work has been restricted to exploring the use of machine 
learning methods for regression, a computationally attractive and well established technique. Features are extracted from normalized NMR spectra, generated from ab initio calculation and simulation based on model structures; output values correspond to unit cell parameters used in computations. 
Model structures are generated by randomly permuting unit cell parameters, rejecting those candidates whose interatomic distances violate steric considerations. In other words, should bond lengths be less than the sum of excepted ionic 
radii, candidates are rejected. 

This work examines both Multiple-Input, Multiple-Output Support Vector Regression (MSVR) and Artificial Neural Networks (ANNs), 
comparing computational time, scaling and accuracy of methods, when used to discern the mapping between input spectra and unit cell parameters of materials that give rise to spectra.
When presented with the simulated spectra for a related polymorph of the structures used during the training phase, both methods reproduce most unit cell parameters fairly accurately.  The overall approach  should be amenable 
to other form of materials data, which in conjunction with suitably constructed and arranged machine learning elements would comprise an expert system for solid state materials characterization, for the elucidation of complex materials.

\section{Theory}
\subsection{Density Functional Theory}

The following serves only as a brief review of DFT, which approximates the many body Schrodinger equation for $N$ electrons, in terms of an interaction between a single electron (with 
wavefunction $\psi_i$ and energy $\epsilon_i$) and a charge density :  

\beq \left[ -\frac{\hbar^2}{2m}\nabla^2 + V(\textbf{r})\right] \psi_i(\textbf{r}) = \epsilon_i \psi_i(\textbf{r}) \eeq

where the charge density $n(\textbf{r})$ is given by :

\beq n(\textbf{r}) = \sum_i^N |\psi_i(\textbf{r})|^2 \eeq

The effective potential comprises three terms :

\beq V(\textbf{r}) = V_{ext}(\textbf{r}) + V_{ee}(\textbf{r}) + V_{xc}(\textbf{r }) \eeq

where $V_{ext}(\textbf{r})$ is the external potential, and $V_{ee}(\textbf{r})$ and $V_{xc}(\textbf{r})$ are the electron-electron repulsion and exchange-correlation contributions, both 
functionals of the density. A variety of approximations have been devised over the years for the latter, two common approaches are the Local Density Approximation (LDA) and the Generalized 
Gradient Approximation (GGA). The solution to equation 1 is obtained via a self-consistent iterative process, whereby approximations to ground-state wave functions are produced after 
convergence. In extended solids, the most popular basis for the expansion of these states is plane-waves that in conjunction with equation 1 gives rise to a generalized eigenvalue problem. The 
excessive number of plane waves required in this expansion for non-valence electronic states prompted the creation of pseudopotentials, where the rapidly oscillating wavefunction near the core 
is replaced by pseudized, smoothed approximations with fewer nodes.  

\subsection{Nuclear Magnetic Resonance}

Once ground-state wavefunctions have been deduced, one may calculate measurable quantities for the material, for example, NMR parameters. The most significant interaction in NMR is between the 
nuclear magnetic moment and an applied static magnetic field (Zeeman effect), producing the Larmor precision at frequency $\omega_0$, which is observable at radio frequencies~\cite{sli}. In 
both liquid and solid state NMR, shifts to these frequencies are produced by the interaction between induced electronic currents and nuclear magnetic moment, the chemical shift interaction. The 
dipole interaction between magnetic moment of different nuclei produces significant line broadening, reduced to a large degree in the liquid state by the tumbling motion of molecules. Indeed, 
rapid interpretation of liquid state NMR spectra has been a routine analytic technique in organic chemistry for many decades. In liquid state NMR, lines in the frequency spectrum are generally 
well resolved and chemical shifts are directly correlated with bond lengths and chemical species, particularly for organic materials.  In terms of magnitude, the most significant interaction 
besides the chemical shift for relevant nuclei is the quadrupole interaction, a function of the electric field gradient $V$ at the nucleus, whose components contribute to the measurable 
quadrupole coupling constant $C_Q$ and asymmetry parameter $\eta$:

\beq C_Q = \frac{eQV_{zz}}{h}; \eta = \frac{V_{yy}-V_{xx}}{V_{zz}} \eeq

Only nuclei with spin $I > 1/2$ including isotopes $^{47}$Ti ($I$ = 5/2) and $^{49}$Ti ($I$ = 7/2) have a non-zero quadrupole moment $Q$, which couples with the electric field gradient. This 
introduces anisotropic line broadening to the NMR spectra of powdered solids composed of many crystallite orientations $\alpha,\beta$ distributed over the unit sphere. Materials are frequently 
studied in this form, given the difficulty of synthesizing single crystal examples. The associated line broadening greatly complicates interpretation; lines for distinct chemical sites overlap 
and simulation is necessary in order to extract parameters such as $C_Q$ and $\eta$ in the solid state. Magic Angle Spinning (MAS) is an experimental technique that alleviates broadening, reducing or removing first order effects; to second order, average Hamiltonian theory gives for the quadrupole frequencies~\cite{man}:

\[ \omega_{r,c} = -\frac{r-c}{\omega_0}\left[ \frac{C_Q}{2I(2I+1)} \right]^2 \left\{ A^{(0)}(I,r,c)\left(\frac{\eta^2+3}{10}\right) +\right. \]

\beq \left. A^{(4)}(I,r,c)f(\eta,\alpha,\beta)\right\} \eeq

These frequencies are a function of the energetic transition $r\leftrightarrow c$, crystallite orientation $\alpha,\beta$, spin $I$ and aforementioned quadrupole parameters. For the purposes of this work, 
attention is restricted to this interaction. Figure 1 provides two examples of simulations using equation 5, for the central frequency transition ($r,c=1/2,-1/2$) measured in common experiments.

\begin{figure}   
\begin{center}
\includegraphics[width=0.9\textwidth]{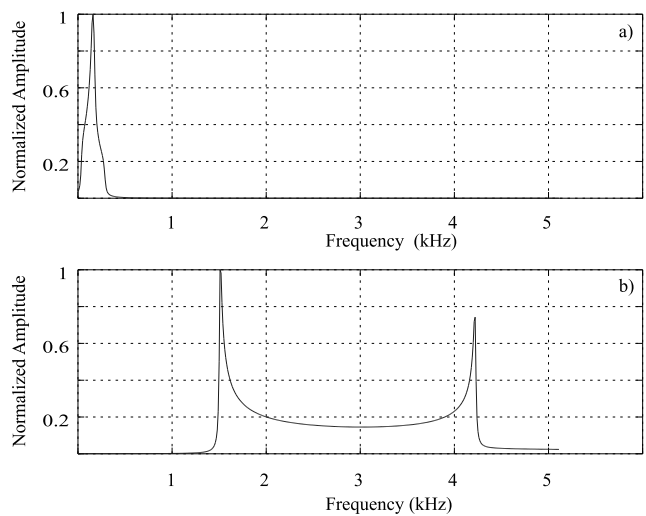} \caption{Simulated $^{47}$Ti MAS lineshapes for a) a material with quadrupole coupling constant of 1MHz and assymetry parameter of 1.0 and b) a material with quadrupole coupling constant of 5Mhz and assymetry parameter of 0, Larmor frequency of 50.75 MHz and 5.12 kHz bandwidth. } 
\end{center}
\end{figure} 

\subsection{Support Vector Regression}
The application of neural networks to regression is well documented, particularly in finance~\cite{dun,ref}, however the adaptation of the support vector approach to regression is more recent.
The linear regression problem is commonly stated as solving for ${\bf w,b}$ in :
\beq {\bf y}=f({\bf x}) = \langle {\bf w,x} \rangle +{\bf b} \eeq
where $x_i,y_i$ are input-output data pairs. The solution process minimizes the norm, under the assumption that the optimal solution approximates all data pairs with precision $\epsilon$. It may be shown that a solution to the optimization problem leads to a Lagrangian formulation, whose solution for the weights ${\bf w}$ is a linear function of input data, the support vector expansion~\cite{smo,bur}.
In general, the relationship between input (feature) and output (value) space is non-linear, and kernels $\phi$ are applied in order to map features to a higher dimensional space, in order to maintain a solution form comprising a linear combination of support vectors and features $\phi({\bf x})$. Support vector regression has been generalized further to solve multiple-input multiple-output (MIMO) problems, abbreviated in this work as MSVR~\cite{msvr}. In this technique, the following Lagrangian expression is minimized:
\beq L_P({\bf W,b}) = \frac{1}{2}\sum^M_{j=1}||{{\bf w}^j}^2|| + C\sum^n_{i=1}L(u_i) \eeq 

where $C$ is a constant analogous to the soft margin parameter, {\bf W,b} are now multi-dimensional regression parameters to be determined, 
\beq {\bf W} = [{\bf w}^1,...,{\bf w}^M]; {\bf b}=[{\bf b}^1,...,{\bf b}^M]^T \eeq

and $L(u_i)$ is defined as :
\beq L(u_i) = \left \{ \begin{array}{ll} 0, & u_i < \epsilon \\ 
u_i^2-2u_i\epsilon+\epsilon^2, & u_i \geq \epsilon \\ \end{array} \right . \eeq
Equation 9 is an expression of the penalty for predictions lying outside the desired precision $\epsilon$, a function of feature-value combinations $\{{\bf x}_i,{\bf y}_i\}$ and transformation kernels ${\bf \phi}$ as follows:
\[ u_i = ||{\bf e}_i|| = \sqrt{{\bf e}_i^T{\bf e}_i} \]
\beq {\bf e}_i^T = {\bf y}_i^T - {\bf \phi}^T({\bf x}_i){\bf W}-{\bf b}^T \eeq
The training set consists of $i=0,...,n$ examples, and ${\bf x}$,${\bf y}$ have dimensions $N,M$ respectively. The scaler $u_i$ is distinct for each input example, the norm of vector ${\bf e}_i$, in turn a function of output ${\bf y}_i$, features ${\phi({\bf x}_i)}$ and regression parameters. The minimization procedure is a non-linear problem solvable in an iterative fashion, and therefore an expression for the gradient is required. This is developed from the aforementioned equation using a Taylor expansion, ultimately leading to the solution of a linear system, for each step of the iterative procedure. After regression parameters have been deduced, one may develop predictions for new input {\bf x}':
\beq {\bf y}' = \phi^T({\bf x}') \Phi^T {\bf \beta}, \eeq
using an expansion for weights as a function of training input {\bf x} and parameters $\beta^j$ optimized by the iterative procedure :
\beq {\bf w}^j = \sum_i \phi({\bf x}_i)\beta^j = \Phi^T \beta^j, \eeq
the multiple-output analog of the support vector expansion for single-output data.

\section{Experiments}
\subsection{Methods}

As alluded to previously, the main goal of this work is to determine the efficacy of machine learning methods in deriving structural details directly from input solid state NMR spectra. Therefore, both DFT computations from model structures, and corresponding NMR simulations are required, in order to create input features (from spectra) and output values (unit cell parameters used in DFT calculations).  
Titania (titanium oxide) was chosen as the material for this work, owing to it's industrial and environmental relevance and wealth of available published information~\cite{jim,downs}, table 1.
Programs from 
within the Quantum Espresso~\cite{qe} suite were used for the pseudopotential (ld1.x), DFT (pw.x) and NMR parameter calculations (gipaw.x). Simulations  of the measurable $^{47}$Ti NMR spectrum 
were performed using custom software~\cite{bro}, and the method detailed in the previous section was  coded in C++ for the machine learning (regression) steps, using support vectors.  Before proceeding with the batch process to be outlined shortly, 
pseudopotentials for Ti and O were generated using the Perdew-Burke-Ernzerhof~\cite{pbe} exchange-correlation functional, capable of calculating magnetic response (NMR) parameters 
using the Gauge Including Projector Augmented Wave (GIPAW)~\cite{gipaw} Method.  All DFT calculations were performed using a mesh of $4\times4\times4$ $k$-points in order to sample the 
Brillouin zone, ultimately providing values of quadrupole coupling constant and asymmetry parameter for rutile and anatase in good agreement with experimental values~\cite{nmr}. The following 
batch process was performed, beginning with the structure for Anatase :
\begin{enumerate}
\item Generate TiO$_2$ structure with fixed Ti coordinates and angles throughout, independently perturb O fractional coordinates $x\equiv y,z$ and unit cell parameters $a\equiv b,c$, with random displacements.
\item If new atomic positions violate steric considerations (ie., distance between Ti$^{4+}$ and O$^{2-}$ is less than sum of ionic radii = 2 A), reject the move, else:
\item Calculate ground state electronic structure for system via DFT
\item Calculate $^{47}$Ti quadrupole coupling constants and asymmetry parameters
\item Simulate $^{47}$Ti MAS NMR spectrum composed of 512 amplitude points, using a Larmor frequency of 50.75 MHz and 5.12 kHz bandwidth.
\end{enumerate}

Therefore, the output value space is of dimension 4 (cell parameters $O_x\equiv O_y,O_z,a\equiv b,c$) and input feature space is of dimension 512. Principle Components Analysis (PCA) was used in this work to compress the input data space, in order to expedite the training process. Referring to the input data as rectangular matrix $X$, with $n$ rows corresponding to different experiments and $N=512$ columns, PCA proceeds by first finding the eigenvalues of $X'X$. This is generally accomplished by computing the more tractable singular value decomposition :
\beq X = U\Sigma W' \eeq
where the non-zero elements of $\Sigma$, the singular values, correspond to the square roots of the eigenvalues $\lambda$ of $X'X$ and $W'$ corresponds to the eigenvectors of $X'X$. By retaining the $N'$ largest eigenvalues and corresponding eigenvectors in columns of matrix $P$, $X$ may be transformed to a smaller space $n\times N'$ using the transformation $Y=X*P$. 

\begin{table}[h]
\begin{center}
\caption{Key polymorphs of titania; $\alpha,\beta,\gamma=90$, Ti=(0,0,0)}
\begin{tabular}{lllccccc}
Formula & Name & Group & $a,b$ & $c$ & $O_x,O_y$ & $O_z$  \\
\hline
TiO$_2$ & Rutile &  P4$_2$/mnm & 4.5922 & 2.9574 & 0.30496 & 0  \\
TiO$_2$ & Anatase &  I4$_1$/amd & 3.7842 & 9.5146 & 0 & 0.20806  \\
\end{tabular}
\end{center}
\end{table}

\subsection{Results}
By generating data after the fashion described, a region within a four dimensional output parameter space is effectively explored. This example benefits from the relatively high symmetry of titania 
and associated reduction in computation time. Over 1000 experiments as described prior were performed using eight Intel Sandy Bridge processors, in under twelve hours. 
A dataset was extracted from experiments, with inputs and outputs as described, by thresholding and selecting the first 500 elements with quadrupole coupling constant $\leq$ -0.95 MHz. The initial dataset contained both positive and negative values for $C_Q$, and measurable second order quadrupole frequencies are insensitive to the sign of $C_Q$. Figure 2 displays the NMR parameters for the selected dataset.

\begin{figure}   
\begin{center}
\includegraphics[width=0.8\textwidth]{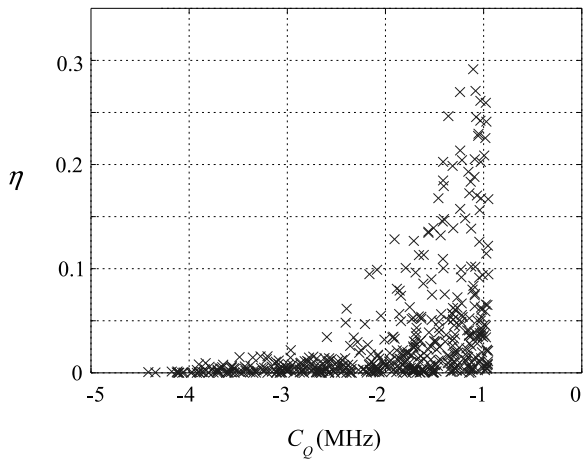} \caption{NMR parameters for dataset; structures with larger magnitude $C_Q$ and smaller $\eta$ are similar to anatase, while  structures with smaller magnitude $C_Q$ and larger $\eta$ are similar to rutile.} 
\end{center}
\end{figure}

After thresholding, the effects of compression on training time and accuracy were investigated. Figure 3 displays a comparison between training time for the nnet package within R, used to implement an ANN, and the aforementioned MSVR method, as implemented in C++. Referring to this figure, as expected, MSVR scales almost linearly with input dimension, while the ANN scales poorly. Training time for the latter increases exponentially with input data dimension, and at input dimension of 128, exceeded the limitations of the package (1000 weights). In order to asses accuracy during experiments, the Root Mean Squared Error (RMSE) was used :
\beq \Delta = \sum_i^n \sqrt{\frac{(y'_{i}-y_{i})^2}{n}} \eeq
where $y_i'$ are values predicted by the particular machine learning method employed, $y_i$ are the real values, and the sum is carried out over the number of test examples $n$ in the datafold. 

\begin{figure}   
\begin{center}
\includegraphics[width=0.8\textwidth]{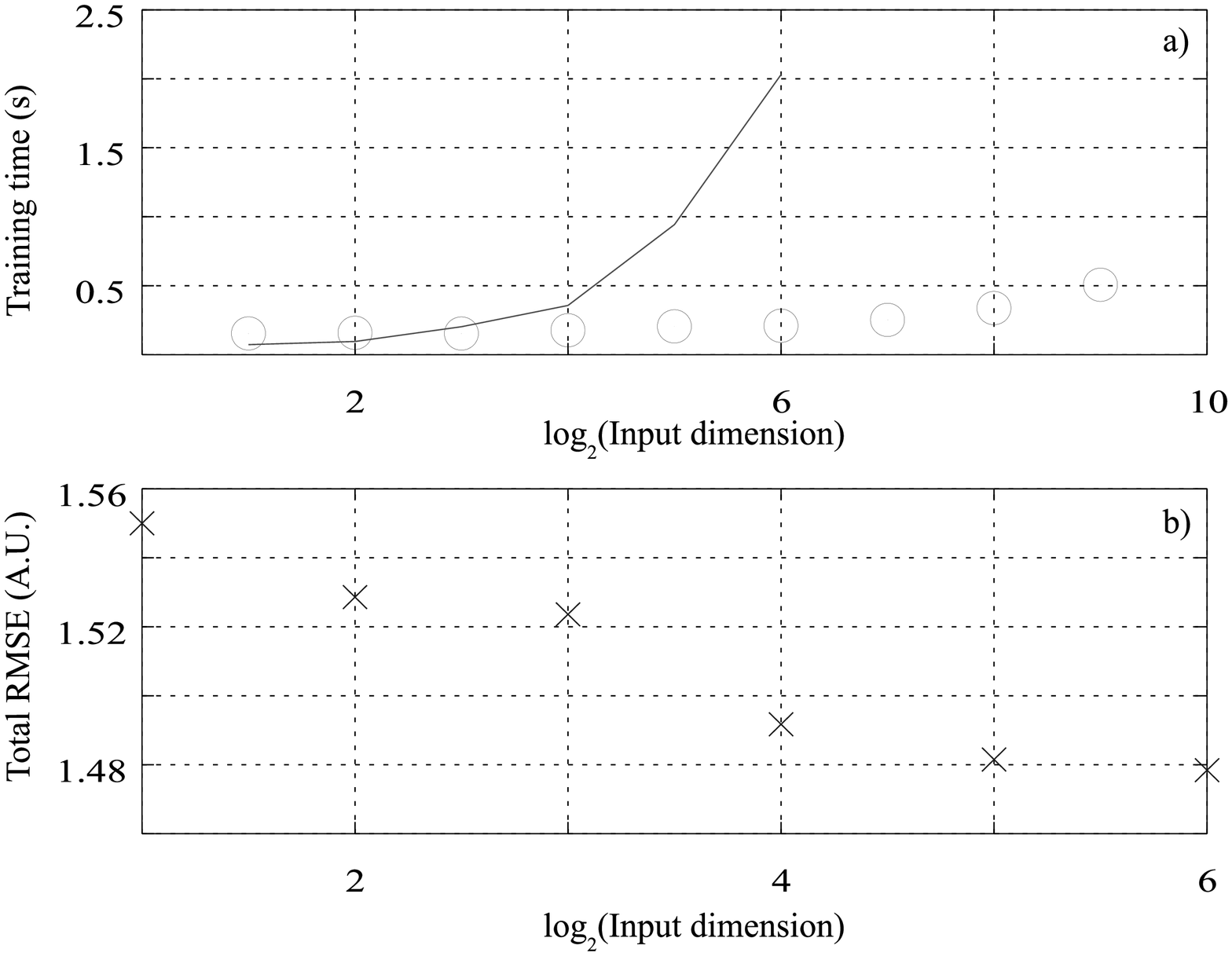} \caption{Results of scaling studies : a) time required to train the ANN (solid line) and MSVR (circles) in seconds, versus the log of the input dimension b) the total RMSE of the ANN as a function of the log of the input dimension. } 
\end{center}
\end{figure}

It was observed that both methods maintain a high degree of accuracy with compression (figure 3b), an encouraging result with regards to establishing a practical database. The difference in total RMSE for both methods was approximately 5\%, between using an input dimension of 2 and 64 features. This is due in no small part to the simplicity of the system and thus spectra; in a more practical situation spectra are much more feature rich, for example in the presence of multiple, overlapping spectral lines (inequivalent chemical sites). For the remainder of experiments, the input data dimension was compressed to 4 input features.
Both methods were compared using ten-fold cross validation, in order to assess overall accuracy and costs of parameter optimization. The RMSE data for both methods is recorded in tables 2 and 3; clearly both methods have comparable error. However, the ANN for small input data dimensions is optimized fairly quickly (56 weights, 4-4-4 network) in distinction to MSVR, which at this stage requires tuning of $C$ and kernel function parameters for a desired precision, in addition to the optimization procedure for regression variables ${\bf W,b}$. With regards to MSVR, the only kernel functions that produced reasonable output were radial basis functions (with optimal $\gamma$ in the range 0.01 to 0.1); linear and polynomial kernels produced far more inaccurate predictions.
Overall, these machine learning methods do show that solid-state NMR provides a sensitive measure for certain unit cell parameter displacements. Figure 4 shows the distributions of relative error $|1-y_i'/y_i|$ from all data folds (10$\times 50=500$ test samples). Fractional coordinate $O_z$, parameters $a\equiv b$ and $c$ are predicted to within less than 10\% relative error, 51\%, 93\% and 44\% of the time respectively, while predictions for $O_x\equiv O_y$ are wildly inaccurate in many instances. The fractional coordinates for O in $x,y$ dimensions were observed to have less than 10\% relative error in less than 13\% of instances. This points to limitations in using a single NMR interaction and indeed single spectroscopic data source in making accurate predictions for all unit cell parameters. A larger expert system would of course incorporate more NMR interactions (including chemical shift) and other data sources, for instance X-ray spectra.

\begin{figure}   
\begin{center}
\includegraphics[width=0.9\textwidth]{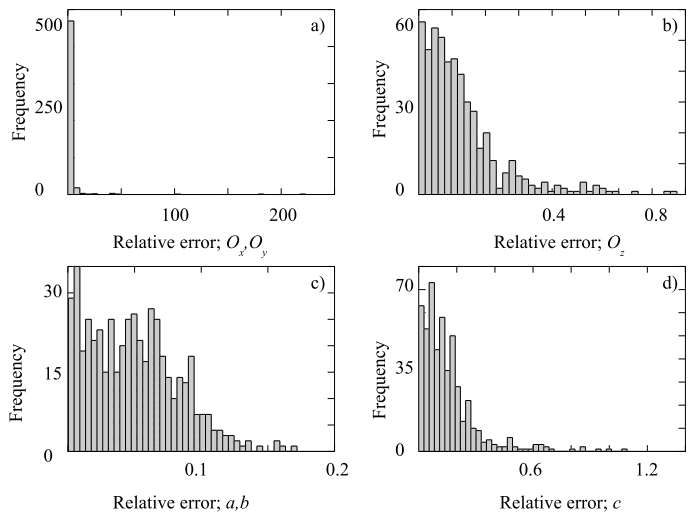} \caption{Distribution of relative errors using MSVR prediction : a) $O_x,O_y$ fractional coordinate b) $O_z$ fractional coordinate c) unit cell parameter $a,b$ and d) unit cell parameter $c$.} 
\end{center}
\end{figure}

\begin{table}[h]
\begin{center}
\caption{MSVR results; RMSE for ten data folds (450 train, 50 test elements, $\epsilon$=0.2)}
\begin{tabular}{lllll}
$\Delta(O_x\equiv O_y)$ & $\Delta(O_z)$ & $\Delta(a\equiv b)$ & $\Delta(c)$ & $C$ \\
\hline
   0.063600 &  0.045500 &  0.243000&   1.055900&   5.600000\\   
   0.073900 &  0.045000 &  0.239500&   1.461700&   6.000000\\   
   0.077300 &  0.054400 &  0.240800&   1.396600&   4.600000\\  
   0.062000 &  0.040200 &  0.252300&   1.255500&   5.000000\\  
   0.066600 &  0.043100 &  0.224400&   1.147100&   6.000000\\   
   0.055800 &  0.045000 &  0.236700&   1.389000&   6.000000\\   
   0.079200 &  0.045600 &  0.268100&   1.479500&   5.000000\\   
   0.060100 &  0.057300 &  0.241400&   1.231500&   3.200000\\   
   0.072600 &  0.045500 &  0.303200&   1.245300&   2.600000\\   
   0.082800 &  0.049300 &  0.227800&   1.261700&   3.000000\\  
\hline
\end{tabular}
\end{center}
\end{table}

\begin{table}[h]
\begin{center}
\caption{ANN results; RMSE for ten data folds (450 train, 50 test elements)}
\begin{tabular}{llll}
$\Delta(O_x\equiv O_y)$ & $\Delta(O_z)$ & $\Delta(a\equiv b)$ & $\Delta(c)$  \\
\hline
 
   0.061488  & 0.039988&   0.239628&   1.187439\\
   0.069671  & 0.044444&   0.219666&   1.403309\\
   0.070845  & 0.053725&   0.236521&   1.500705\\
   0.061350  & 0.038591&   0.237223&   1.303657\\
   0.067755  & 0.042339&   0.219838&   1.172234\\
   0.057470  & 0.043959&   0.221471&   1.314616\\
   0.074059  & 0.046857&   0.228791&   1.488244\\
   0.066621  & 0.049576&   0.235042&   1.190761\\
   0.072513  & 0.039145&   0.228004&   1.339722\\
   0.079725  & 0.045298&   0.225725&   1.436712\\

\hline
\end{tabular}
\end{center}
\end{table}

\section{Conclusions}

This work details several aspects of building a larger expert system for solid state physics, demonstrating the use of MSVR and ANNs in learning the mapping between spectra and structure for model 
systems. These systems are generated using fixed composition but variable atomic positions. The virtue of this approach is that more complicated materials, for example oxide surfaces, are 
composed of simpler systems albeit with unknown atomic coordinates. By repeating the process outlined here for more models, the aim is to create a database and expert system for the elucidation 
of materials including oxide surfaces, composed of simpler systems.  

In a complete expert system for solid state materials, machine learning elements may be trained on various types of 
spectroscopic data for known structures, so that spectra for new materials may be presented to the system and underlying structure deduced rapidly. The contribution presented here is predicated 
on knowledge of the underlying chemical objects comprising the material. In the absence of this knowledge, a process of classification or unsupervised learning must take place first. Also, more 
complex systems including solid solutions generally require large super cells in order to accurately perform DFT calculations of measurable parameters, exponentially increasing the 
dimensionality of the feature space that must be explored, in order to produce reliable models. However, while not addressed in this work, many candidate structures can potentially be ruled out 
on energetic and other grounds. Before a DFT calculation for a condensed system can proceed, pseudopotentials for constituent atoms must exist, constructed using the same exchange-correlation functional to be 
applied to the extended system under study. As mentioned, a pseudopotential drastically reduces the number of plane waves required in a calculation, by using an approximation to the core region 
of the potential experienced by electrons. Considerations as to the suitable partition of valence and core orbitals for a given environment strongly dictates the success or lack thereof in 
using pseudopotentials. As a rule of thumb, explicitly including more valence electrons provides greater transferability to different bonding environments, at the expense of computation time. 

In order to build an expert system, a survey and compilation of appropriate pseudopotentials would need to be performed, with particular emphasis on the ability to reproduce measurable 
quantities such as those used in this work. Finally, parameters such as quadrupole coupling constants in NMR are proportional to tensor traces ie., are insensitive to the sign on atomic 
displacements. Nonetheless, augmented twith information from other iterations in NMR including the chemical shift (particularly sensitive to chemical identity), or other spectroscopic data, 
these limitations are readily overcome. 

\section{Acknowledgement}
This work used resources on the TACC Stampede cluster, available via the Extreme Science and Engineering Discovery Environment (XSEDE) initiative, which is supported by National Science Foundation grant number ACI-1053575.

\end{document}